\documentclass[aps,pre,twocolumn,showpacs,superscriptaddress]{revtex4-2} 
\usepackage{graphicx}  
\usepackage{dcolumn} 
\usepackage{bm}
\usepackage{amssymb} 
\usepackage{amsmath}
\usepackage{dsfont}  
\usepackage[english]{babel}
\usepackage{pgfplots}
\usepackage{textcomp}
\usepackage{color}
\usepackage{soul} 
\begin{document}
\widetext
\title{Nanoparticle Taylor dispersion near charged surfaces with an open boundary}
\author{Alexandre~Vilquin}
\thanks{The authors contributed equally}
\affiliation{Gulliver UMR 7083 CNRS, PSL Research University, ESPCI Paris, 10 rue Vauquelin, 75005 Paris, France}
\affiliation{IPGG, 6 rue Jean-Calvin, 75005 Paris, France}
\author{Vincent~Bertin}
\thanks{The authors contributed equally}
\affiliation{Gulliver UMR 7083 CNRS, PSL Research University, ESPCI Paris, 10 rue Vauquelin, 75005 Paris, France}
\affiliation{Univ. Bordeaux, CNRS, LOMA, UMR 5798, F-33405, Talence, France}
\affiliation{Physics of Fluids Group, Faculty of Science and Technology, and Mesa+ Institute, University of Twente, 7500AE Enschede, The Netherlands.}
\author{Elie Rapha\"el}
\affiliation{Gulliver UMR 7083 CNRS, PSL Research University, ESPCI Paris, 10 rue Vauquelin, 75005 Paris, France}
\author{David~S. Dean}
\affiliation{Univ. Bordeaux, CNRS, LOMA, UMR 5798, F-33405, Talence, France}
\affiliation{Team MONC, INRIA Bordeaux Sud Ouest, CNRS UMR 5251, Bordeaux INP, Univ. Bordeaux, F-33400, Talence, France.}
\author{Thomas~Salez}
\email[]{thomas.salez@cnrs.fr}
\affiliation{Univ. Bordeaux, CNRS, LOMA, UMR 5798, F-33405, Talence, France}
\author{Joshua~D.~McGraw}
\email[]{joshua.mcgraw@cnrs.fr}
\affiliation{Gulliver UMR 7083 CNRS, PSL Research University, ESPCI Paris, 10 rue Vauquelin, 75005 Paris, France}
\affiliation{IPGG, 6 rue Jean-Calvin, 75005 Paris, France}
\date{\today}
\begin{abstract}
The dispersive spreading of microscopic particles in shear flows is influenced both by advection and thermal motion. At the nanoscale, interactions between such particles and their confining boundaries become unavoidable. We address the roles of electrostatic repulsion and absorption on the spatial distribution and dispersion of charged nanoparticles in near-surface shear flows, observed under evanescent illumination. The electrostatic repulsion between particles and the lower charged surface is tuned by varying electrolyte concentrations. Particles leaving the field of vision can be neglected from further analysis, such that the experimental ensemble is equivalent to that of Taylor dispersion with absorption. These two ingredients modify the particle distribution, deviating strongly from the Gibbs-Boltzmann one at the nanoscale studied here. The overall effect is to restrain the accessible space available to particles, leading to a striking, ten-fold reduction in the spreading dynamics as compared to the non-interacting case. 
\end{abstract}
\maketitle

\begin{figure}[t!]
	\includegraphics[scale = 1.0,trim=0 0 0 0,clip=true]{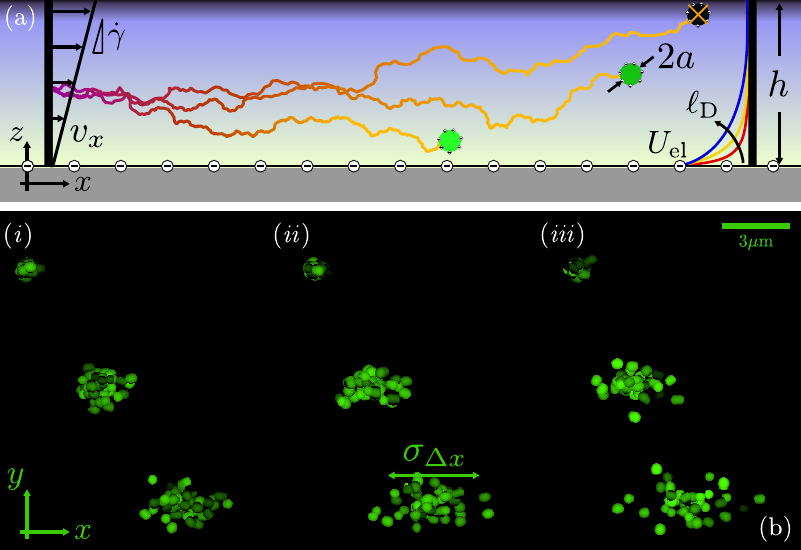}
	\caption{\label{fig:SchemRawData} (a) Side-view schematic of the experimental setup. In a channel of height $h$, nanoparticles with radius $a$ are advected by a linear shear flow, $v_x(z) = \dot{\gamma}z$, and diffuse. Electrostatic potentials $U_\mathrm{el}(z)$ with Debye lengths $\ell_\mathrm{D}$ repel particles from the bottom surface at $z=0$. Particles reaching the upper limit at $z=h$ are considered as absorbed. (b) Reconstructions of three successive experimental positions of fluorescent nanoparticles, with top-, middle- and bottom-row lag times $\tau=\{2.5, 25, 50\}$\ ms, in (\emph{i}) pure water, (\emph{ii}) 5.4 mg/L,  and (\emph{iii}) 54 mg/L NaCl aqueous solution, see SI Videos 1 and~2.}
\end{figure}

Diffusion is a fundamental microscopic transport mechanism that can be effectively enhanced by orders of magnitude in the presence of a hydrodynamic velocity gradient. In the process commonly known as Taylor dispersion~\cite{Taylor1953}, particles starting from the same position are advected along the flow while the concentration profile is broadened due to Brownian diffusion across streamlines, \emph{cf.} Fig.~\ref{fig:SchemRawData}(a). Such enhanced broadening, \emph{i.e.} dispersion, is the principal mechanism for solute dispersal in many natural and technological contexts~\cite{Brenner1993, Grotberg1994, Shapiro1986, Stein2006, Nielsen2010, Tan2012, Marbach2016, Dehkharghani2019, Aminian2016}. At nanometric distances from surfaces, however, particles can no longer be considered simple tracers since they are subject to: intermolecular forces~\cite{Isrealachvili2011}; mobility-reducing hydrodynamic interactions with boundaries~\cite{Brenner1961,Faucheux1994,Prieve1999,Lavaud2021}; as well as reaction/absorption at the latter~\cite{barton1984asymptotic, Shapiro1986, Biswas2007, alexandre2021stickiness}. Such interactions modify the spatial structure of particle-probability distributions, but there is yet no observation about how this modification could affect the diffusive-like transport dynamics of Taylor dispersion. The object of this Letter is thus to link nanoscale probabilistic structure to spreading dynamics. 

Taylor dispersion has many applications in situations where such physico-chemical interactions are important, with biophysical ones as emblematic examples. As such, the seminal theoretical work of Taylor has been significantly extended~\cite{brenner1977constrained, Dill1982, Shapiro1986, Biswas2007, Allaire2010, Marbach2018, Marbach2019, Alonso2019, Peng2020, Kumar2021, Alexandre2021}. Particle absorption or chemical reactions on the boundaries induces a gradual and substantial loss of particle number. This particle loss may strongly modify particle probability distributions, also complexifying models for dispersion~\cite{Marbach2019, Alexandre2021,alexandre2021stickiness}. Reactivity is critical for several applications in chemistry and life science~\cite{Bello1994, Cottet2007, Lee2008, Cottet2010, Ibrahim2013, Chamieh2018, Hong2020, Deleanu2021, Chazot2022}. 

Several experimental works noted that such interactions or absorption~\cite{Pedro1993} bias diffusion coefficient evaluations using the Taylor dispersion method~\cite{Bello1994}. In typically millimetric capillary tubes, the induced error was noted as a few tens of percent~\cite{Madras1996}. Such errors may become more significant in smaller systems, for instance in applications for peptide diffusion and the determination of aggregate sizes~\cite{hawe2011taylor,Deleanu2021}, among others~\cite{moser2021taylor}. State-of-the-art Taylor dispersion studies also focused on the geometry of a flow domain~\cite{Aminian2016} at micro- and milli-metric scales. Other recent experiments, reaching micro- and nano-scales~\cite{Cuenca2012} focussed on pre-asymptotic dispersion dynamics~\cite{Fridjonsson2014,Takikawa2019,Vilquin2021}, but not on the link between nanoscale statistical distributions and Taylor dispersion.

In this Letter, we study nanoscale Taylor dispersion in a near-surface shear flow. First, we systematically vary the role of surface interactions by tuning the repulsive electrostatic interaction between the nanoparticles and the lower surface, becoming important for dispersion when the corresponding interaction scale is comparable to that of confinement. Second, we employ a finite observation zone with an open upper boundary, particles leaving this zone formally correspond to permanently absorbed ones. Our experiments thus allow a study of nanoscale dispersion under adsorption, without the obvious inconvenience of a polluted, physically absorbing surface. Such a particle loss is found to strongly modify the observed particle probability distribution relevant for dispersion. Using an extended moment theory~\cite{Aris1956, Barton1983, Camacho1993, Vedel2012, Vedel2014}, we quantitatively recover our experimental observations. In contrast to the few tens of percent noted above for macroscopic systems, we observe a ten-fold Taylor dispersion reduction when both boundary effects are present, as compared to the case when they are absent.

We used objective-based, total internal reflection fluorescence microscopy (TIRFM)~\cite{Fish2009, Vilquin2021}, observing individual, negatively charged, $a = 55$\ nm-radius, latex colloidal particles. The particles were suspended in pressure-driven shear flows near an interface between salted water and glass (see Supplementary Material \S I and \S II at Ref.~\footnote{Please see Supplementary materials (SM) at [URL] for: raw and reconstructed data videos; further details concerning the experiments; unscaled dispersion data as in Fig.~2(a); and details of the modelling, these latter details which are referenced throughout the text. The SM also includes the Refs.~\cite{novotny_principles_2006,bailey2001sleign2}} for videos and experimental details). These observations yield the particle positions (the apparent height $z_\mathrm{app}$ and $x,y$ in-plane positions), resolved to within a few tens of nanometers. The former allows access to altitude probability distributions (APDs), whereas sequential particle observations were linked into temporal trajectories (\emph{ca.} $10^5$ of them for this study), see SM Video~1. Displacements $\Delta x =  x(t+\tau)-x(t)$ over a delay time $\tau$ from a particle's first observation time $t$ were thus recorded and used to determine the near-wall shear rate, $\dot{\gamma}$ (see SM  \S II). Also obtained were the variances of the stream-wise, $\sigma_{\Delta x}^2 = \langle (\Delta x -\langle\Delta x\rangle )^2\rangle$ (see Fig.~\ref{fig:SchemRawData}(b)), and transverse displacements, $\sigma_{\Delta y}^2$, allowing to calculate dispersion and diffusion coefficients. Independent trajectories were superimposed at common spatio-temporal origins, as in Fig.~\ref{fig:SchemRawData}(b) and SM Video 2, to visualise the evolution of particle ensembles. Particle volume fractions were small enough to ignore inter-particle interactions. 

\begin{figure}[t!]
	\includegraphics[scale = 1.0,trim=0 0 0 1mm,clip=true]{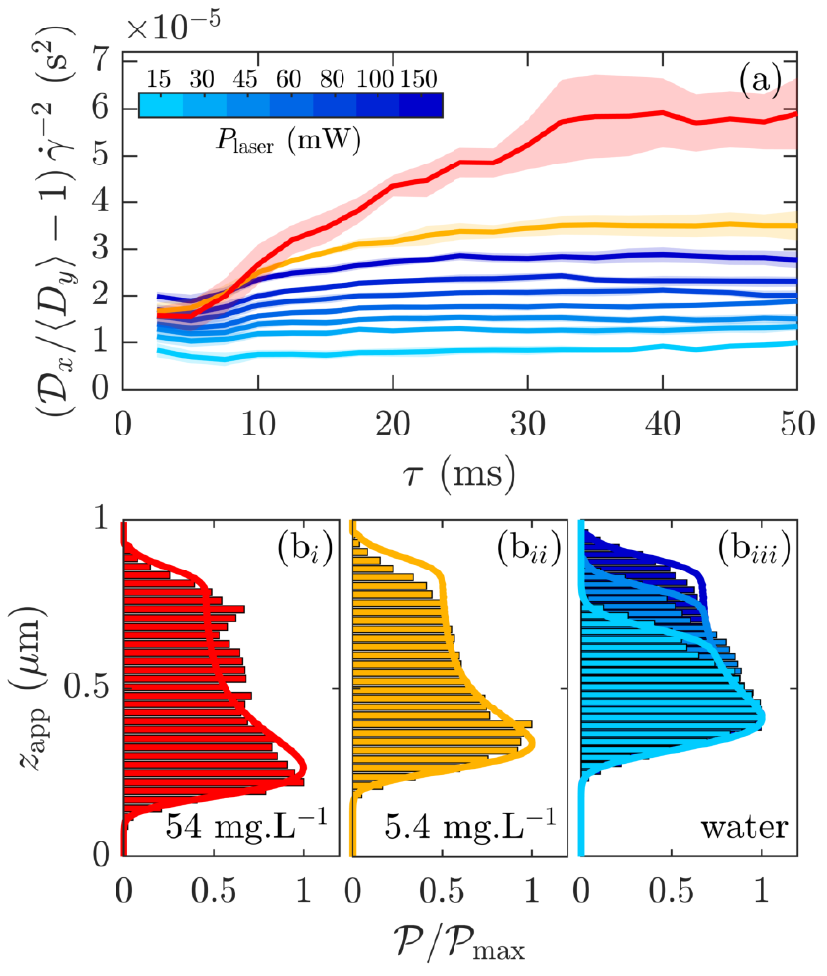}
	\caption{\label{fig:APD-Dispersion} (a) Time dependence of the scaled, shear-rate normalised, dispersion coefficient. Each line corresponds to averaging at least four shear rates, the shaded area displaying the associated standard deviation. (b) Normalized QE-APDs as a function of distance, $z_\mathrm{app}$, for varying salt concentrations: (\emph{i}) $\mathrm{[NaCl]} = 54$ mg/L, (\emph{ii}) $\mathrm{[NaCl]} = 5.4$ mg/L, and (\emph{iii}) ultrapure water. In ($\mathrm{b}_{iii}$), the three laser powers are shown. The same color code is used in (a). }
\end{figure}

Taylor-Aris theory predicts a rescaled, long-time dispersion coefficient  $\mathcal{D}_x/D_0-1 = \mathrm{Pe}^2/30$ for a linear shear flow bounded by reflecting walls~\cite{Barton1983, Vilquin2021}. Here, $\mathcal{D}_x = \sigma_{\Delta x}^2/(2\tau)$, $D_0$ is the bulk diffusion coefficient of the nanoparticles, and $\mathrm{Pe} = \dot{\gamma}h^2/(2D_0)$ is the Peclet number comparing transport by advection and diffusion, $h$ being the observation zone height. In Fig.~\ref{fig:APD-Dispersion}(a) are shown the normalised streamwise dispersion coefficients as a function of $\tau$, with at least four shear rates used for each condition (see SM \S II-D for unscaled data). The normalisation uses the depth-averaged $D_y = \sigma_{\Delta y}^2/(2\tau)$ (note the angle brackets in the axis label), which closely approximates the bulk diffusion coefficent $D_0$ \cite{Vilquin2021}. Importantly, we note a strong modification in the dispersion coefficient on changing the salt concentration: the data for the highest salt concentration gives nearly a three-fold increase in the dispersion, as compared to ultrapure water, as also indicated in Fig.~\ref{fig:SchemRawData}(b). We note furthermore that $h$ and $D_0$ are identical for the three different data sets at laser illumination power of $150$ mW in Fig.~\ref{fig:APD-Dispersion}(a). Therefore, the classical Taylor-Aris theory, supposing non-interacting tracer particles in flows bounded by rigid walls, is clearly inappropriate here. This observation motivates a detailed investigation into the influence of the interactions with the walls  on dispersion. 

At equilibrium, the particles' concentration follows a Gibbs-Boltzmann distribution with $c_{\textrm{B}}\propto\exp\left[-U/(kT\right)]$, where $U$ is an interaction potential and $kT$ the thermal energy. In Fig.~\ref{fig:APD-Dispersion}(b$_{\mathrm{\emph{i} - \emph{iii}}}$) are thus shown experimental APDs, $\mathcal{P}$  for identically imposed pressure drops of 30 mbar and different salinities; the distributions are normalised by their maxima and no filtering concerning the time of observation is made. Since the particles and glass surfaces are negatively charged, a repulsive electrostatic interaction that can be obtained from DLVO theory~\cite{Isrealachvili2011} is expected:
\begin{align}
	U_\mathrm{el}(z) &= kT\frac{a}{\tilde{\ell}_\mathrm{B}}\exp\left(-\frac{z-a}{\ell_\mathrm{D}}\right)\ .
	\label{eq:Debye}
\end{align} 
Here $\ell_\mathrm{D}$ is the Debye length, $\tilde{\ell}_\mathrm{B} =e^2/(\epsilon kT)$ $\times\left[\tanh(e\psi_\mathrm{p}/(4kT))\tanh(e\psi_\mathrm{w}/(4kT))\right]^{-1}$ is a surface-modified Bjerrum length~\cite{Isrealachvili2011}, and $e$, $\epsilon$, $\psi_\mathrm{p}$, and $\psi_\mathrm{w}$ are the elementary charge, the dielectric permittivity of the liquid, the particle and wall surface potentials. 

The lines in Fig.~\ref{fig:APD-Dispersion}(b) are model fits to the experimental APDs particularly including the Boltzmann distribution $c_\mathrm{B}(z)$ with the potential of Eq.~\eqref{eq:Debye} as the only energetic contribution --- other necessary ingredients, see~\cite{Zheng2018, Li2015, Vilquin2021}, include: the finite camera sensitivity giving the observation zone height, $h$; and, objective optics and particle polydispersity, modifying the direct correspondence between distance and intensity. The good agreement for the full fits here suggests that the electrostatic repulsion mainly~\footnote{In the fitting, we have only varied the Debye length for the three salt concentrations, as well as the base intensity $I_0$ and the maximum altitude accessed by the particles for the three laser powers.} determines the distribution of particles near the wall. Therefore, we call these the quasi-equilibrium APDs (QE-APD). Quantitatively, the Debye lengths obtained from the QE-APD fits are $\ell_\mathrm{D} = \{67, 32, 10\}\pm3$\ nm for $\mathrm{[NaCl]} = \{0, 5.4, 54\}$\,mg/L, respectively, in agreement with the DLVO theory~\cite{Isrealachvili2011}. Furthermore, we find a salinity-independent $\tilde{\ell}_\mathrm{B} = 13\pm3$ nm consistent with expected particle and wall potentials of approximately $-120$ mV, see SM \S II.A. For the different salinities, the decreased electrostatic repulsion allows particles to access to a larger part of the velocity gradient, enhancing the dispersion. For an observation zone height of approximately 800 nm, a 50 nm Debye-length modification gives a two-fold change in the dispersion coefficient. 

Besides intermolecular interactions, the height $h$ is a key ingredient for the QE-APD fits. This height can be tuned by changing the laser power, as shown in Fig.~\ref{fig:APD-Dispersion}(b$_{iii}$). Accordingly, for pure water, decreasing the laser power gives a further factor of 2 decrease in the normalised, steady $\mathcal{D}_x$ between the highest and lowest laser powers in Fig.~\ref{fig:APD-Dispersion}(a). On exceeding $h$ a particle's trajectory is no longer considered, as indicated by the crossed-out particle in Fig.~\ref{fig:SchemRawData}(a), and thus the open boundary acts as an ideal particle sink. This sink progressively modifies the structure of the particle distribution in the observation zone~\cite{Madras1996,Shapiro1986}, as shown next. 

\begin{figure}[t!]
	\includegraphics[scale = 1.0,trim=0 0 0 0,clip=true]{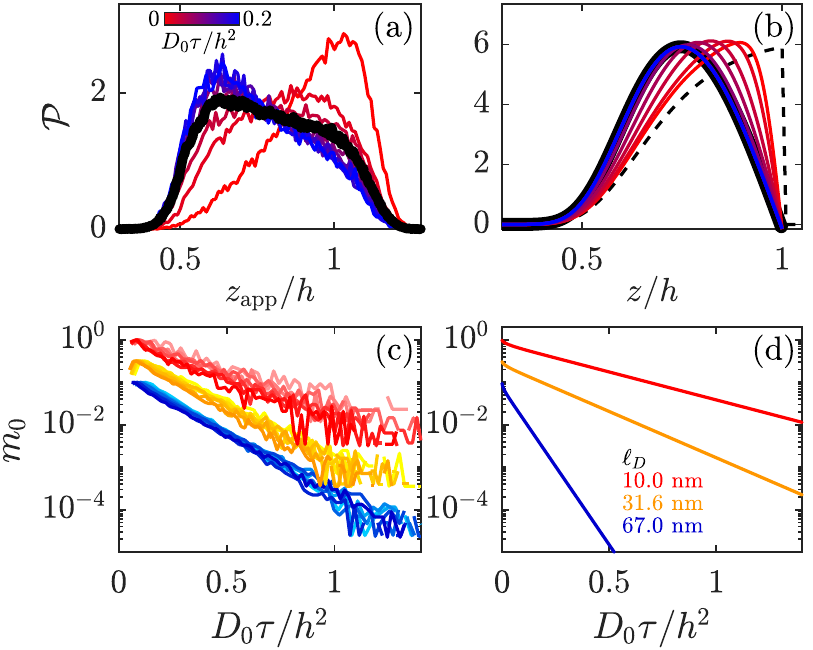}
	\caption{\label{fig:ZerothMoment} (a) Rescaled experimental APDs, for the indicated dimensionless lag times and for ultrapure water, $P_\mathrm{laser} = 150$ mW and a pressure drop of 30 mbar. The black curve shows the QE-APD (\emph{cf.} Fig.~\ref{fig:APD-Dispersion}($\mathrm{b}_{iii}$)). (b) Theoretical prediction for (a), with $3\times10^{-4}\leq D_0\tau/h^2\leq 3\times10^{-1}$. The initial condition (dashed line) corresponds to $c_{\textrm{B}} \propto \exp[-U_\mathrm{el}/(kT)]$, with Eq.~\eqref{eq:Debye} and the parameters obtained through fitting in Fig.~\ref{fig:APD-Dispersion}(b). (c) Experimental and (d) theoretical remaining particle fractions, as functions of dimensionless lag time. The color codes are the same as in Fig.~\ref{fig:APD-Dispersion}, and the shades indicate the same varying laser powers as for water. Curves of different salinity are shifted vertically for clarity.}
\end{figure}

In Fig.~\ref{fig:ZerothMoment}(a), experimental \emph{time-dependent} (TD-)APDs are shown for pure water, displaying different delay times since the particles' first observation. As the typical time scale to diffuse out of the observation zone is given by $h^2/D_0$, the TD-APDs are plotted for different values of the dimensionless time $D_0\tau/h^2$. A temporal evolution of the TD-APD is observed, and a steady-state is reached for times approaching the diffusion time $h^2/\left(D_0\pi^2\right)$ predicted by Taylor~\cite{Taylor1953}. Remarkably, this long-time, steady distribution is different from the QE-APD, thus representing a violation of the Gibbs-Boltzmann distribution, shown in black for comparison.  

To assess the effect on the aforementioned probabilistic modifications on the dispersion, theoretically we consider a population of nanoparticles initially located at the origin $x=0$ (see Fig.~\ref{fig:SchemRawData}) and distributed vertically with an initial concentration profile $c(x=0,z,t=0)=c_\mathrm{ini}(z)\delta(x)$. The concentration field $c(x,z,t)$ obeys the advection diffusion equation~\cite{brenner1977constrained} 
\begin{equation}
\label{eq:Fokker-Planck}
\frac{\partial c}{\partial t} + v_x(z) \frac{\partial c}{\partial x} = D_x(z) \frac{\partial^2 c }{\partial x^2} + \frac{\partial}{\partial z} \bigg( D_z(z)\bigg[ \frac{\partial c}{\partial z} + \frac{U_\mathrm{el}'(z)}{kT} c\bigg] \bigg)\, ,
\end{equation}
where $D_x$ and $D_z$ are the streamwise and cross-stream diffusion coefficients. These latter depend on $z$ due to hydrodynamic forces induced by  the no-slip boundary condition at the hard wall (see SM Eq. (S3)). Zero particle flux at the wall, imposes $D_z\left[ \frac{\partial c}{\partial z} + \frac{U_\mathrm{el}'(z)}{kT} c\right]=0$ at $z=a$. As nanoparticles are not followed after they leave the observation zone, the concentration field vanishes at the open boundary, \textit{i.e.} $c(x,z,t)=0$ at $z=h$. This Dirichlet boundary condition is equivalent to a chemical absorption reaction with an infinite reaction rate~\cite{Biswas2007}. 

The moments of the concentration field described by Eq.~\eqref{eq:Fokker-Planck} can be computed in many ways, including: the moment~\cite{Aris1956,Barton1983,Camacho1993,Vedel2012,Vedel2014}, invariant manifold~\cite{mercer1994complete,Marbach2019}, Green-Kubo~\cite{van1990taylor,Alexandre2021}, and large-deviation methods~\cite{haynes2014dispersion,kahlen2017large}. Here, we use a moment theory involving time-dependent streamwise $p^\mathrm{th}$ (with $p\geq 0$) moments $c_{p}(z,t) = \int_\mathbb{R} x^p c(x,z,t)\,\mathrm{d}x$ recursively (see SM \S III). Using a modal decomposition, the solution is found to be of the form: 
\begin{equation}
	c_p(z,t) = \sum_{k=1}^\infty c_{p,k}(z,t) \exp(-\lambda_k t) \ , 
	\label{EQ:eigendecomp}
\end{equation}
where $c_{p,k}$ are polynomial functions of $t$ of degree $p$, and $\lambda_k$ are the eigenvalues of the corresponding Sturm-Liouville problem, with $\lambda_1 < \lambda_2 < \cdots$. We show in Fig.~\ref{fig:ZerothMoment}(b) the theoretical TD-APD at different times for an ensemble of particles initially distributed according to a Boltzmann weight $c_\mathrm{ini}(z)=c_\mathrm{B}(z)$, using the same electrostatic parameters and the absorbing wall at $h$ obtained by fitting the data in Fig.~\ref{fig:APD-Dispersion}(b). The main qualitative features of the experimental observations are recovered: a depletion zone develops near the open boundary while the TD-APD converges towards a steady distribution, corresponding to the spatial structure of the slowest eigenmode with $\lambda = \lambda_1$ (see SM Eq.~(S9)).

Importantly, the slowest eigenmode of Eq.~(\ref{EQ:eigendecomp}) has a nonzero eigenvalue such that the total number of particles $m_0$ decays exponentially at long times. This decay is a consequence of the absorbing boundary at the limit of the observation zone. In Fig.~\ref{fig:ZerothMoment}(c), we show the experimental fraction $m_0(t) = \int_a^h c_0(z,t) \, \mathrm{d}z / \left[\int_a^h c_0(z,0) \, \mathrm{d}z\right]$ of particles remaining in the observation zone, as a function of the dimensionless lag time. No matter the strength of the electrostatic interactions and the laser power, a temporally exponential decay of the number of particles is observed at long times. Similarly, in Fig.~\ref{fig:ZerothMoment}(d), we show the theoretical fraction of particles remaining in the observation zone, as a function of the dimensionless lag time, for the three Debye lengths accessed accessed experimentally. We again find an exponential decay at large times, \emph{i.e.} $m_0\propto\exp\left(-\lambda_1\tau\right)$, independent of $c_\mathrm{ini}(z)$.

\begin{figure}[b!]
	\includegraphics[scale = 1.0,trim=0 0 0 0,clip=true]{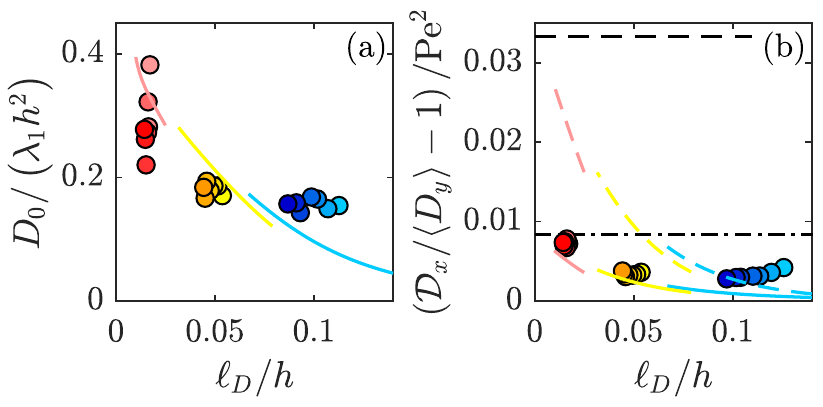}
	\caption{\label{fig:TimesDispersions} (a) Dimensionless decay time of the number of remaining particles vs. height-normalized Debye length; colors as in Fig.~\ref{fig:ZerothMoment}. Theoretical predictions with absorbing boundary conditions are displayed using solid lines (see SI). (b) Reduced, steady dispersion coefficient as a function of the height-normalized Debye length. Theoretical predictions are described in the text, solid lines using the same theory as in (a).  SM Figure S3 schematically describes each model. }
\end{figure}

From a microscopic point of view, the nanoparticles diffuse out of the observation zone, such that the typical decay time scale is set by the time $\sim h^2/D_0$ needed for the particle to reach the absorbing boundary at the top of the observation zone. Besides, the decay time depends on the electrostatic and hydrodynamic interactions via the ratios between the typical length scales in the problem and the channel size. Altogether, the theoretical decay rate reads $\lambda_1 = \frac{D_0}{h^2} F\left(\frac{\ell_\mathrm{D}}{h}, \frac{\tilde{\ell}_\mathrm{B}}{h}, \frac{a}{h}\right)$, where $F$ is an unknown dimensionless function to be determined by solving the eigenvalue problem described in the SM \S III-B. In Fig.~\ref{fig:TimesDispersions}(a), we compare experiments and theory for the dimensionless decay time $D_0/\left(\lambda_1 h^2\right)$ as a function of the Debye-length-to-channel-size ratio. As expected, the longer the range of the electrostatic interaction (\textit{i.e.} the larger Debye length), the faster particles leave the observation zone. While there is a small deviation between the measurements and predictions, especially for the unmodified water (blue), the overall trends agree.

Since the open boundary of our experiments affects the TD-APDs, as described in Figs.~\ref{fig:ZerothMoment}(a,b) and as dictated by the spatial structure of the slowest eigenmode, our theoretical approach allows a prediction regarding dispersion. Computing the first and second moments of the concentration, we extract the dispersion coefficient of the remaining particles, see SM \S III-E. This coefficient  converges to a steady value at long times, as in Fig.~\ref{fig:APD-Dispersion}(a). The long-term dispersion coefficient $\mathcal{D}_x$ can be written as the sum of the steady-state averaged streamwise molecular diffusion coefficient $\langle D_x \rangle$, \emph{cf.} SM Eq.~(S40), and a term induced by the advection-diffusion coupling:
\begin{align}
	\mathcal{D}_x = \langle D_x \rangle + \int_a^h\frac{1}{c_{\mathrm{B}}(z)}\left[v_x(z)-\langle V \rangle\right]\zeta_1(z)f_1(z)\ \mathrm{d}z\ ,
	\label{eq:MomentDisp}
\end{align}
where $\langle V \rangle $ and $f_1(z)$ are the steady-state averaged velocity, and the steady TD-APD shown in Fig.~\ref{fig:ZerothMoment}(b), respectively. The quantity $\zeta_1$ is an auxiliary function related to $f_1(z)$ as in SM Eq.~(S19). 

In Fig.~\ref{fig:TimesDispersions}(b) are shown the rescaled, steady dispersion coefficients for all of the experimental salinities and laser powers (colored dots). The general increase of dispersion coefficient with salinity seen is expected due to increased access to the near-wall regions on electrostatic screening, as in Fig.~\ref{fig:APD-Dispersion}. For a quantitative description of the data, we also display predictions of four different~models (see SM Fig.~S3 for schematics). 

The tracer theory of Taylor and Aris~\cite{Taylor1953, Aris1956} largely over-estimates the data (black dashed line). Moment theory for a wall with infinite adsorption rate, \emph{i.e.} an open boundary, but no lower surface interactions with the wall (SM Eq.~(S42); black, dashed-dot line) predicts a significant global decrease in the dispersion coefficient. Both these models are yet independent of salt concentration. The theory of Refs.~\cite{brenner1977constrained, Alexandre2021} assumes a reflective boundary at $z=h$ and includes conservative interactions with one wall. Dispersion coefficients from this theory read $\mathcal{D}_x = \langle D_x\rangle+f[D_z(z), c_\mathrm{B}(z), v_x(z)]$, \emph{cf.} SM Eq.~(S44), and are decreased as compared to the classical Taylor model (dashed, colored lines), yet still overestimate the measured dispersion coefficients. 

Finally, the moment theory combining electrostatic interactions and an open boundary at $z=h$, Eq.~(\ref{eq:MomentDisp}), quantitatively captures the measurements (solid colored lines), even while noting a small systematic deviation for the pure water case. Fig. 4(b) stresses that using an absorbing boundary condition at the limit of the observation zone is necessary to accurately estimate the reduction of the dispersion coefficient measured in the TIRFM experiments. Indeed, the relevant statistical distributions at the heart of Taylor dispersion phenomena are thereby strongly modified. In contrast, the existing theory~\cite{brenner1977constrained, Alexandre2021} depends only on the APD given by $c_\mathrm{B}(z)$.

To conclude, our collective observations show that chemical- or absorption-induced leakage at boundaries and particle-surface interactions can play a dominant role in Taylor dispersion at the nanoscale. Nanoscale transport is routinely used to measure the physical properties of biological objects and processes, as in Refs.~\cite{hawe2011taylor, jelinska2017denaturation, Hong2020, Deleanu2021, moser2021taylor, Chazot2022}. Here, we have demonstrated that the precise nature of particle-surface interaction must be carefully taken into account to yield accurate measurements.

\begin{acknowledgments}
\noindent The authors gratefully acknowledge Pierre Soulard and Fr\'ed\'eric Restagno for fruitful and ongoing discussions, as well as Caroline Cramail and Aym\`ene Sadaoui for preliminary experimental implications. This work has also benefited from the technical contribution of the joint service unit CNRS UAR 3750.  The authors benefited from the financial support of CNRS, ESPCI Paris, the Agence Nationale de la Recherche (ANR) under CoPinS (ANR-19-CE06-0021), EMetBrown (ANR-21-ERCC-0010-01), Softer (ANR-21-CE06-0029) and and Fricolas (ANR-21-CE06-0039) grants, and of the Institut Pierre-Gilles de Gennes (Equipex ANR-10-EQPX-34 and Labex ANR-10-LABX- 31), and PSL Research University (Idex ANR-10-IDEX-0001-02). They also thank the Soft Matter Collaborative Research Unit, Frontier Research Center for Advanced Material and Life Science, Faculty of Advanced Life Science at Hokkaido University, Sapporo, Japan. Finally, the authors acknowledge financial support from the European Union through the European Research Council under EMetBrown (ERC-CoG-101039103) grant. Views and opinions expressed are however those of the authors only and do not necessarily reflect those of the European Union or the European Research Council. Neither the European Union nor the granting authority can be held responsible for them. 
\end{acknowledgments}


%

\end{document}